# The Angular Momentum of Stars Reflects the Relationship between Star-forming Environment and Galactic Evolution History

Yu-Fu Shen (申淯夫)[1], Yan Xu (许妍)[1,2], Yi-Bo Wang (王夷博)[1], Xiu-Lin Huang (黄修林)[1], Xing-Xing Hu (胡星星)[1], and Qi Yuan (袁琦)[1]
[1] Changchun Observatory, National Astronomical Observatories, Chinese Academy of Sciences, People's Republic of China; shenyf@cho.ac.cn
[2] School of Astronomy and Space Sciences, University of Chinese Academy of Sciences, People's Republic of China
*Received 2025 January 19; revised 2025 June 5; accepted 2025 June 5; published 2025 July 18*

## Abstract

This study focuses on stars with masses above the Kraft break in the Kepler field. Their rotational angular momenta are essentially the same as those at the zero-age main sequence. The angular momentum dissipation experienced by these stars during their pre-main sequence (PMS) phase is also relatively weak, so their rotational angular momentum can reflect the parameters of their parental molecular clouds. The reliability of angular momentum measurements was evaluated based on the phenomenon of angular momentum conservation observed in stars before and after the turnoff point in observational data. We find that stars with masses between $1.4M_\odot$ and $1.8M_\odot$ show an inverse proportionality between angular momentum and isochrone age. We propose that the angular momentum–age correlation reflects changes in the star-forming environment in the Milky Way's history. Besides, the observed inverse proportionality implies that as the Milky Way has evolved, the stars formed within it tend to possess greater rotational angular momenta. This trend would promote the fragmentation of stars during the PMS phase and inhibit the formation of massive stars, providing a useful perspective for explaining variations in the initial mass function.

*Unified Astronomy Thesaurus concepts:* Galaxy evolution (594); Stellar kinematics (1608); Stellar rotation (1629)

*Materials only available in the online version of record:* machine-readable table

## 1. Introduction

The observations of A. A. Goodman et al. (1993) have given rise to the angular momentum puzzle in the process of star formation: If all the angular momentum of the molecular cloud is inherited by the stars, the rotation speed of the stars would be much higher than the breakup speed, making it impossible for them to maintain stability. However, the rotation speed of stars typically remains below the breakup speed, with one of the main spin-down mechanisms being the star–disk interaction during the pre-main sequence (PMS; J. Ferreira et al. 2000; S. Matt & R. E. Pudritz 2005; C. Zanni & J. Ferreira 2009; S. P. Matt et al. 2010; C. Zanni & J. Ferreira 2013; F. Gallet & J. Bouvier 2015; Y. Bu et al. 2025). In addition, gravitational torques also prevent a star from spinning faster than approximately 50% of its breakup speed during formation (M.-K. Lin et al. 2011). Lower-mass stars are more strongly affected by the spin-down mechanisms, with speeds roughly at 10% of the breakup speed (L. Hartmann & J. R. Stauffer 1989; W. Herbst et al. 2007), while higher-mass stars have speeds that exceed 20–40% (S. C. Wolff et al. 2006; W. Huang et al. 2010), partly due to the disk-clearing accretion phase significantly reducing the stellar angular momentum, and massive stars are more difficult to spin-down in the disk-clearing phase due to their larger inertia and weaker magnetic fields (A. L. Rosen et al. 2012). Another significant spin-down phase occurs in the main sequence (MS) phase, where low-mass stars rapidly spin-down due to wind losses, resulting in rotation periods longer than about 10 days within 1 Gyr (e.g., A. Skumanich 1972; L. G. Bouma et al. 2023). However, when the stellar mass exceeds a certain threshold, the fraction of angular momentum loss to the total angular momentum becomes negligible. This is because higher-mass stars possess weaker magnetic fields, leading to weaker stellar wind, combined with the effect that higher-mass stars retain more total angular momentum from the PMS phase. The threshold at which wind angular momentum losses can be ignored is known as the Kraft break, typically demarcated at $T_{\rm eff} > 6250$ K. In addition, for stars with even greater mass, such as those exceeding $2M_\odot$, the stellar wind becomes dominated by radiation pressure and significantly intensifies again, but this work hardly involves such stars.

There are two primary methods for measuring stellar rotation periods. Traditionally, spectroscopy was used to measure the stellar rotation via the rotational broadening of absorption lines (e.g., R. P. Kraft 1965, 1967). However, factors such as thermal motion of atoms, turbulence, and mutual collisions between particles in the stellar atmosphere can also broaden the absorption lines, necessitating repeated iterations based on multiple high signal-to-noise ratio absorption lines to balance these parameters. Currently, the most commonly used method is to analyze light curves, especially when attempting to obtain a large sample of stellar rotation periods. The magnetic active regions of stars have lower $T_{\rm eff}$ and rotate along with the star, resulting in periodic variations in the light curve in the optical band, also known as spot modulation.

Compared to rotational velocity, angular momentum is a more physical quantity, and rotational velocity is merely a partial manifestation of a star's angular momentum. When analyzing the rotational angular momentum of the sample, the inclusion of turnoff stars in the sample can provide extremely important insights. As a star exits the MS phase and enters the turnoff phase, its radius and rotational speed change abruptly, while angular momentum is conserved, in the absence of external influences like planets (which may be swallowed during expansion) or binary companions.

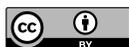

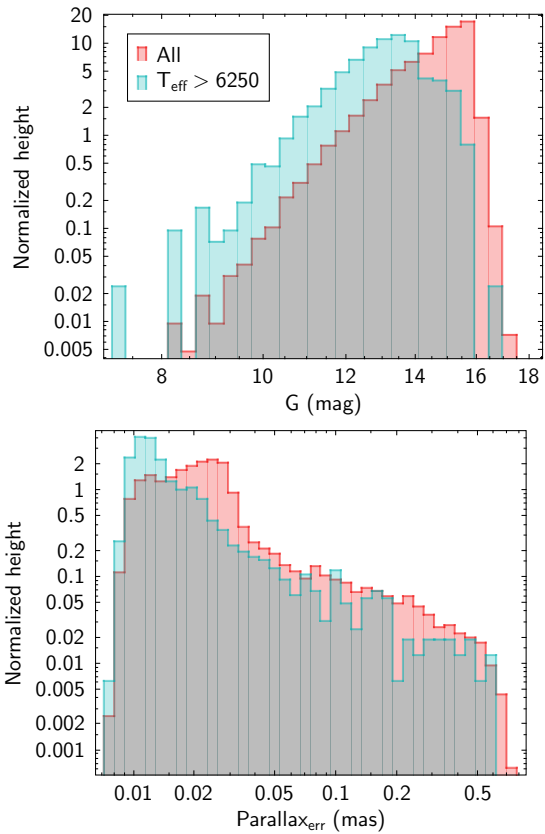

**Figure 1.** The distribution of $G$ magnitude and parallax errors for 31,638 stars, with a subset of 3175 stars having $T_{\rm eff} > 6250$ K.

To observe the conservation of angular momentum, we must ensure that the stellar parameters which have strong influence on angular momentum remain unchanged between the MS and turnoff phases, and then investigate the impact of other parameters on the rotational angular momentum. This implies that the masses of the stars in the sample should be similar, and their formation environments should also be similar, because the formation environment and mass jointly determine the angular momentum of a star at the zero-age main sequence (ZAMS), and mass also affects the angular momentum evolution of the star during the MS phase. This is inherently complex but unexpectedly simple for turnoff stars, because the turnoff stars typically observed have masses greater than approximately 1.2 solar masses, due to the excessively long MS lifetimes of low-mass stars. Most of these turnoff stars happen to be above the Kraft break, which means that MS evolution does not need to be considered.

As a result, stars above the Kraft break retain more information about their angular momentum from their ZAMS phase; similarly, massive stars also experience weaker angular momentum loss in the PMS phase (A. L. Rosen et al. 2012), preserving information about their parental molecular clouds. Therefore, the rotational angular momentum of turnoff stars should reflect the variation in the properties of molecular clouds throughout the formation history of the Milky Way.

In this work, we first discuss the reliability of the sample parameters, including the potential impact of binary star and selection effects. Then, we focus on stellar radius $R$ ($R_\odot$) and rotation period $P_{\rm rot}$ (day) of stars, define $RR_P = R \cdot R/P_{\rm rot}$ as easily accessible proxy for angular momentum, and find the relationship between angular momentum and isochrone age. Subsequently, we consider the potential impact of the moment of inertia, velocity distribution, and interactions between the Milky Way and neighboring dwarf galaxies, ultimately concluding that what is presented in the sample is the angular momentum–age distribution of Galactic field stars. Finally, we discuss the angular momentum evolution during the PMS and MS phases, arguing that this distribution largely reflects the angular momentum of stars at the ZAMS phase, with the spin-down during the MS phase accounting for a small proportion. Finally, we provide a logically closed-loop explanation based on existing observational phenomena and numerical simulations.

## 2. Data

We focused on two sets of stellar parameters. The first is from Gaia's Final Luminosity Age Mass Estimator (FLAME[3]) program, which employs Bayesian inference based on the probability distribution of samples on the HR diagram to derive stellar parameters, using the GSP-Phot parameters[4] as input. These are parts of Gaia DR3 (Gaia Collaboration et al. 2016, 2023; F. van Leeuwen et al. 2022) The second set is from T. A. Berger et al. (2020), which provides stellar parameters based on broadband photometry and Gaia parallaxes (hereinafter referred to as B20), using the isochrone method.

B20 considers broadband photometry, so $T_{\rm eff}$ in this paper are all from B20. Reliable radii can be obtained based on $T_{\rm eff}$ and the high-precision parallaxes and photometry from Gaia. Figure 1 shows the distribution of $G$ magnitudes and the parallax error for the sample. Since this work focuses on higher-mass stars, especially turnoff stars, this subset of the sample is brighter and the parallax parameters have relatively better precision, which makes the radii good. Both FLAME and B20 radii are fine, and we use the B20 radii because, as mentioned later, we use the B20 isochrone ages.

Next, we determine the reliability of the mass and age estimates. For low-mass MS stars, it is almost impossible to distinguish their ages without considering rotation (Z. R. Claytor et al. 2020). The ages given by FLAME for low-mass stars are unreliable; they do not exhibit a rotation-age distribution, as shown in Figure 2. B20 performs better than FLAME in terms of the rotation period-age distribution of low-mass stars (still shown in Figure 2). As for stars above the Kraft break, the measurements from both methods are valid, but B20 is superior, which we have proven using open clusters.

There are a very small number of stars from open clusters in the sample. Among them, only NGC 6819 has an age greater than 1 Gyr, and the rest have ages less than 1 Gyr, which are not of much significance for this work. Age measurements based on stellar parameters for very young stars are also highly inaccurate. Based on T. Cantat-Gaudin & F. Anders (2020), we have identified the NGC 6819 member stars in our sample, three of which have entered the turnoff phase, as shown in Figure 3. The age of this cluster (W. S. Dias et al. 2021) is indicated in the figure. Within the mass range of approximately 1.3–1.5$M_\odot$, the ages derived from FLAME seem to be better than those from B20. However, when calculating the mean age and variance for all member stars, B20 (with a mean age of 2.44 and a variance of 0.65) outperforms FLAME (which has a mean age of 2.81 and a variance of 0.81). Furthermore, FLAME appears to underestimate errors. Therefore, this work exclusively uses the B20 parameters. It should be noted that Figure 3 appears to show an overall decrease in age with increasing mass, which may be due to some prior introduced by the B20 method, or it may be related to a certain age scatter existing within the open cluster itself. Finally, we will categorize the sample by mass for a separate analysis, thus weakening the possible influence of priors from B20.

---
[3] https://gea.esac.esa.int/archive/documentation/GDR3/Data_analysis/chap_cu8par/sec_cu8par_apsis/ssec_cu8par_apsis_flame.html
[4] https://gea.esac.esa.int/archive/documentation/GDR3/Gaia_archive/chap_datamodel/sec_dm_astrophysical_parameter_tables/ssec_dm_astrophysical_parameters.html

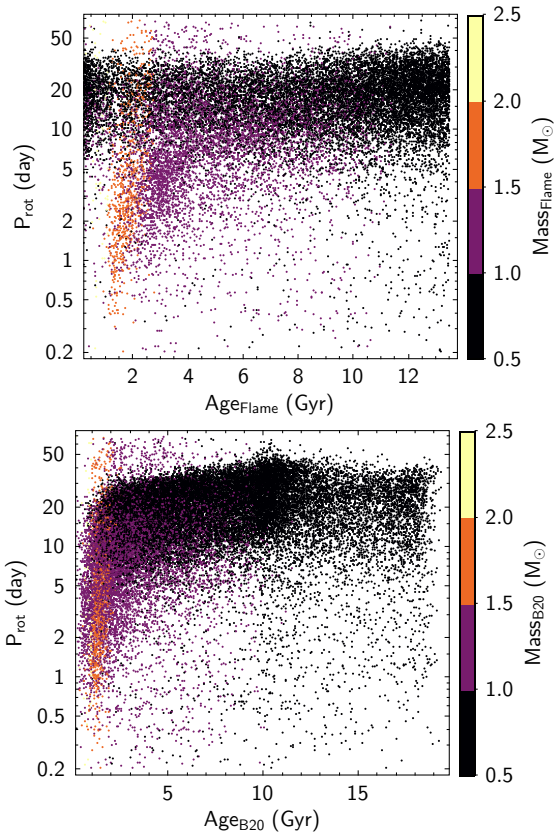

**Figure 2.** Rotation period vs. age of 31,638 stars. The color represents stellar mass.

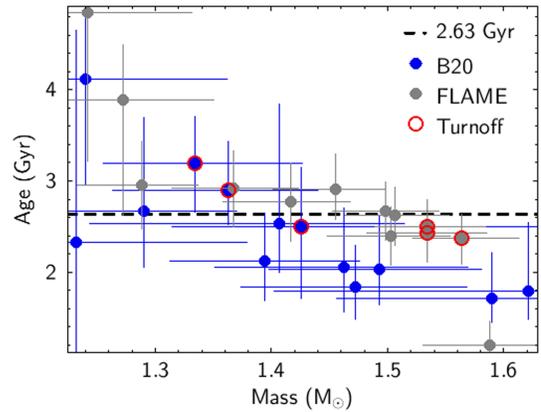

**Figure 3.** Mass vs. age of star members of NGC 6819. Both FLAME ages (gray) and B20 ages (blue) are shown. Turnoff stars are marked by red circles. The horizontal lines represent the age of NGC 6819 given by W. S. Dias et al. (2021): 2.63 Gyr.

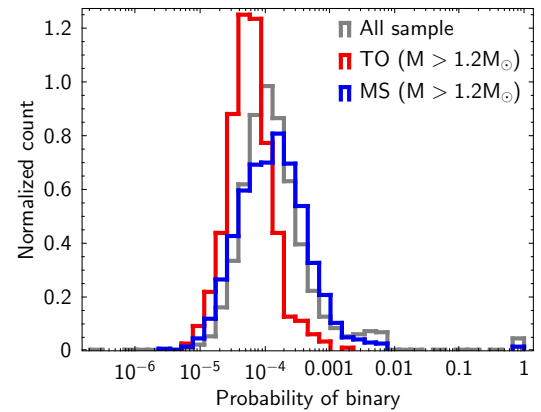

**Figure 4.** The distribution of the probability that an object in the sample is a binary. The red points are $>1.2M_\odot$ turnoff stars; the blue points are $>1.2M_\odot$ MS stars.

Finally, we use the rotation periods from light curves (A. McQuillan et al. 2014). However, the clumpy and time-evolving nature of the underlying active region distribution makes the light curves neither sinusoidal nor strictly periodic, so Fourier domain methods are not always effective. A. McQuillan et al. (2013) used the autocorrelation function (ACF) to address this issue and later applied it to the MS stars (A. McQuillan et al. 2014)[5] in the Kepler sample (W. J. Borucki et al. 2010; D. G. Koch et al. 2010). The 4-year-long light curves provided by Kepler are much longer than the stellar rotation periods, yielding highly confident results. The ACF method can attenuate the influence of spurious peaks from jumps and long-term systematics, and can also bridge the peak splits caused by spot evolution and differential rotation, yielding a clear average rotation period. In addition, pulsation effects do not need to be considered for the MS or turnoff star sample. These series of advantages make the rotation periods given by A. McQuillan et al. (2014) reliable.

In summary, the samples used in this work were obtained by cross-matching Kepler IDs with A. McQuillan et al. (2014) and T. A. Berger et al. (2020), totaling 31,638 stars, with stellar parameters from T. A. Berger et al. (2020), which is based on the cross-match between Kepler and Gaia in T. A. Berger et al. (2018).

### 2.1. Impact of Binarity

Considering that close binary companions cause stars to rotate more rapidly than single stars (E. A. Avallone et al. 2022), which will also affect the angular momentum, consideration of binaries in the sample is necessary. Gaia provides the probability of being a binary star from DSC-Combmod.[6] As seen in Figure 4, the proportion of suspected binaries in the sample is extremely low. As a precaution, we still removed suspected binaries but did not base the removal on the probability from DSC-Combmod, as this value involves complex priors. We chose to retain samples with RUWE (Renormalised Unit Weight Error) < 1.4 (L. Lindegren 2018), totaling 26,513 stars. Therefore, the impact of binaries is not discussed further in this work.

---

[5] It actually contains many turnoff stars whose masses are underestimated due to being misidentified as MS, but this has no impact on the measurement of rotation periods.

[6] https://gea.esac.esa.int/archive/documentation/GDR3/Gaia_archive/chap_datamodel/sec_dm_astrophysical_parameter_tables/ssec_dm_astrophysical_parameters.html

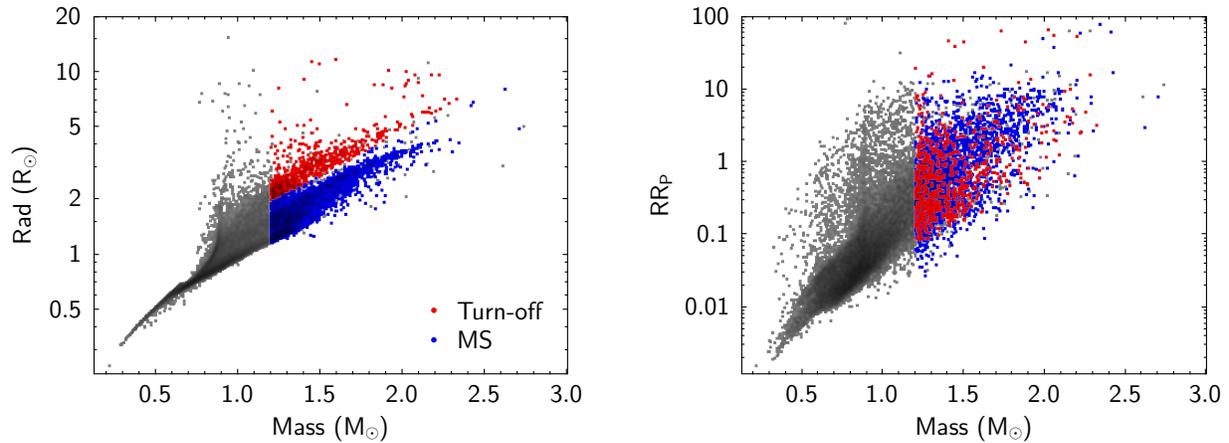

**Figure 5.** Angular momentum conservation after entering the turnoff phase. The red points are turnoff stars in the target sample, the blue points are MS stars in the target sample, and the other stars are gray.

## 3. Analysis

### 3.1. Angular Momentum Conservation and Its Implications

In the sample of this work, we observed the conservation of angular momentum. We focus on the sample with B20 masses greater than $1.2M_\odot$ and RUWE < 1.4, hereinafter referred to as the target sample, which is close to the mass corresponding to the Kraft break. Within the target sample, we further distinguish between MS stars (blue) and turnoff stars (red) using the boundary defined by

$$\log_{10}(R_*) = 0.49 \cdot (M_* - 1.26) + 0.31.$$

This is shown in Figure 5. Among the 3694 stars in the target sample, 712 are turnoff stars and 2982 are MS stars. We classify turnoff stars and MS stars based on their radii and mass, a process that does not involve rotational periods at all. However, the $RR_P$ distribution shows that the distributions of these two types of stars are close, suggesting that angular momentum conservation is observed after entering the turnoff phase. It should be emphasized that the rotational periods of turnoff stars are generally larger than those of MS stars, which is only to be expected. Therefore, it is not sufficient to merely observe that the distribution of turnoff stars is closer to that of MS stars in the $RR_P$ distribution than in the radius distribution. The two must have nearly overlapping $RR_P$ distributions to indicate that angular momentum conservation has been observed. Additionally, the turnoff stars obtained by mass and radius may be mixed with subgiants, but we focus solely on the phenomenon of abrupt radius changes while conserving angular momentum in stars after they leave the main sequence. Therefore, it is unnecessary to distinguish between turnoff and subgiant, and there is currently no strict boundary between them. If one wants to distinguish between the turnoff and the subgiant, a reference can be made to the boundary given in A. B. A. Queiroz et al. (2023).

To understand the physical implications of the observed angular momentum conservation phenomenon, an in-depth analysis is required. The angular momentum of a star depends on two factors: the stellar mass and the stellar formation environment. Regarding Figure 5, the angular momentum increases significantly with mass. Therefore, it is considered that in this sample, the stellar mass has a dominant influence on the angular momentum. Thus stars of the same mass must be observed to see angular momentum conservation; if the mass is measured incorrectly, angular momentum conservation cannot be observed. There is a common situation that can lead to incorrect mass measurement: when one assumes that the entire sample consists of MS stars. Since turnoff stars are cooler than MS stars of the same mass, in this case, turnoff stars are confused with lower-mass MS stars.

The formation environment, as a secondary factor in this sample, is reflected in the scatter in the angular momentum–mass relationship. However, it remains unclear which factors in the stellar formation environment have a greater influence on the angular momentum of stars at the ZAMS. The easily accessible parameters of the stellar formation environment are [Fe/H] and age. Different formation environments may share the same [Fe/H], and the influence of [Fe/H] might be counterbalanced by variations in other elements. Moreover, the same age could correspond to different stellar formation environments. Neither [Fe/H] nor age can fully describe the stellar formation environment, but we can evaluate the extent to which these parameters affect the angular momentum of stars at the ZAMS. By observing the $RR_P$–[Fe/H] and $RR_P$–age relationships, it can be easily seen that the relationship between age and $RR_P$ is more pronounced, as shown in Figure 6. Therefore, this work prioritizes the analysis of the $RR_P$–age relationship and then analyzes whether [Fe/H] has an influence on $RR_P$.

The $RR_P$–age of the target samples exhibits the pattern shown in Figure 6, where the angular momentum appears to decay over time, spanning 2 orders of magnitude. However, before analyzing the specific reasons for this phenomenon in detail, the influence of other factors should be examined.

### 3.2. Selection Effect

It is necessary to determine whether the samples in this work are representative. Longer-period rotation is generally more difficult to detect for two reasons. First, longer-period rotation requires observational data that span a longer period of time, but the Kepler data span up to four years, so the impact of this aspect can be ignored. Second, the magnetic activity of slower rotating stars is weaker, making the variations in their light curves less pronounced. In summary, massive stars exhibit a more severe selection bias. Figure 7 shows the mass distribution of the sample in A. McQuillan et al. (2014) and that of all observed Kepler stars, indicating that a selection effect does indeed exist but the target sample can still represent at least half of the stars within the same mass range in the overall sample. Moreover, even if a large number of slowly rotating massive stars were omitted, they would only be added below the target sample in Figure 6, and the clear trend of $RR_P$–age at the upper side of the distribution would remain unchanged.

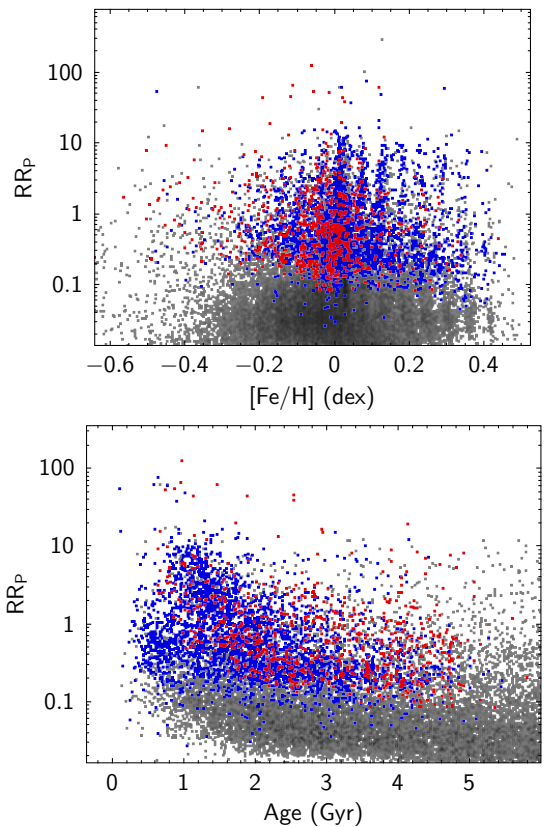

**Figure 6.** The $RR_P$–[Fe/H] and $RR_P$–age distributions. The colors have the same meanings as in Figure 5.

Considering that stars above the Kraft break lack an outer convection zone, the appearance of any sunspots (or starspots) would seem rather peculiar. Therefore, the proportion of cases in the region greater than approximately 1.2 in Figure 7 should be lower. However, starspots are not the only magnetic effect capable of producing rotational modulation. First, circumstellar plasma clumps in magnetically enforced corotation with the star have already been discovered in stars above the Kraft break (e.g., J. D. Landstreet & E. F. Borra 1978; T. Kiker et al. 2024). Second, turbulence could force small-scale magnetic fields (e.g., V. I. Abramenko et al. 2011; M. Stangalini et al. 2025). Theoretically, such small-scale dynamos may occur in stars above the Kraft break, but the specific conditions remain uncertain. Third, fossil magnetic fields (e.g., J. Braithwaite & H. C. Spruit 2004; F. R. N. Schneider et al. 2019) are another possible source of magnetic fields in early-type stars, typically associated with binary systems but not exclusively. Determining which cases in the sample fall into these categories is beyond the scope of this work.

In addition to the selection effects inevitably introduced by the measurement of rotation periods, the Kepler mission also has its own inherent selection effects. Kepler went by brightness to increase S/N and look for late-type main sequence stars. Therefore it naturally ended with F- and G-types (N. M. Batalha et al. 2010). As shown in Figure 8, for stars in the target sample, the MS stars are predominantly of F-type and G-type, while the turnoff stars are mainly G-type with a small number being K-type. Therefore, the overselection of G-type stars is equivalent to overselecting older MS stars and younger turnoff stars. The selection effect on turnoff stars is relatively weak because the majority of turnoff stars are already G-type. Incidentally, it is because Kepler overselected G-type stars that our sample contains a considerable number of turnoff stars. Oversampling within specific age ranges can lead to spurious correlations for the following reason. Suppose that there is no correlation between $RR_P$ and age. However, since $RR_P$ itself contains some significant outliers, since there is no correlation, these outliers are evenly distributed across various age ranges. The overpopulated age ranges will exhibit a higher number of outliers, while the underpopulated age ranges, due to their small sample sizes, will have almost no outliers. When outliers are several orders of magnitude larger than typical $RR_P$, they can shift the average to larger values, creating spurious correlations. The additional selection of G-type stars causes the $RR_P$–age relationship for MS stars to exhibit a tendency toward a positive correlation, while turnoff stars tend to show a negative correlation under the same influence. Since the sample is predominantly composed of MS stars, and the number of turnoff stars is not only small but also less impacted by selection effects, we conclude that on the whole Kepler's selection effects lead to a positive correlation in the $RR_P$–age relationship of the target sample.

### 3.3. Influence of Moment of Inertia and Stellar Velocity Distribution

In reality, $RR_P$ is not equivalent to angular momentum $L$. The relationship is given by $L \propto I/MR^2 \cdot M \cdot RR_P$, where $I$ represents the moment of inertia. For a uniform sphere, the value of $I/MR^2$ is 0.4, whereas for the Sun, it is 0.07.[7] Based on the calculations in A. Claret & A. Gimenez (1989), we performed interpolation on the 3694 sample stars that are the focus of this work and $I/MR^2$ ranges approximately between 0.03 and 0.14. However, we cannot directly multiply $I/MR^2$ by $RR_P$, because for stars near the turnoff point, there is no well-established theoretical model to describe the variations in $I/MR^2$. Consequently, $I/MR^2$ not only carries measurement errors arising from interpolation based on stellar parameters but also introduces systematic biases due to the model's limitations. The parameters provided in A. Claret & A. Gimenez (1989) exhibit high sensitivity to age near the turnoff point. If $I/MR^2$ were multiplied by $RR_P$, the impact of age uncertainties would be further amplified and propagated into $RR_P$, thereby introducing an artificial correlation in the $RR_P$–age relationship. Although the variations in $I/MR^2$ are not substantial enough to significantly alter the numerical value of $RR_P$, this artificial correlation could interfere with correlation analyses. Therefore, instead of multiplying $I/MR^2$ by $RR_P$ and applying error propagation formulas, we opted to amplify the measurement errors of $RR_P$. The degree of amplification was based on the distribution of the interpolated $I/MR^2$ values. For our sample, the mean interpolated value of $I/MR^2$ is 0.046, with a variance of 0.013, resulting in a normalized error of 0.28.

---

[7] https://nssdc.gsfc.nasa.gov/planetary/factsheet/sunfact.html

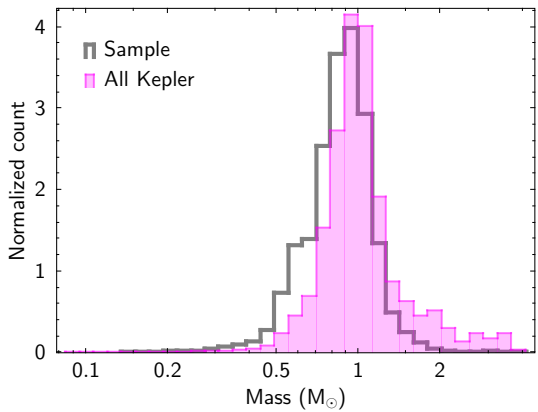

**Figure 7.** The mass distribution of all observed Kepler stars and those with rotation periods. For consistency, the masses in this plot are provided by the Kepler Catalog. This figure is intended to qualitatively demonstrate the selection effects of rotation period measurements.

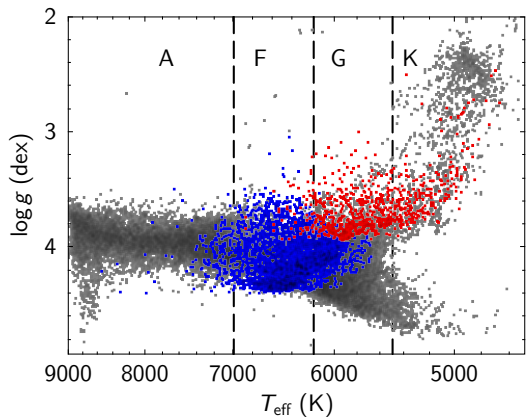

**Figure 8.** The distribution of $T_{\rm eff}$ vs. $\log g$ for samples in LAMOST DR10 with signal-to-noise ratios greater than 100 in all of the *ugriz* bands (as represented by the gray dots). The three vertical dashed lines approximately delineate the boundaries for A-, F-, G-, and K-type stars, with reference to the spectral type classification provided by LAMOST. The distribution of the target sample is superimposed on the LAMOST distribution (gray) to display their corresponding spectral types, and the colors have the same meanings as in Figure 5.

In addition, stars also exhibit differential rotation. The rotation rates provided by Kepler data are obtained by detecting quasiperiodic brightness variations, which arise as magnetically active regions on the star's surface rotate in and out of view. The resulting rotation period should be close to the period at the latitude where sunspots most frequently appear, which is around 16° for the Sun. The maximum difference in the surface rotation period of the Sun is approximately a factor of 1.4 (R. Howe et al. 2000). In addition, the velocity distribution within the star also differs. When deeper than the tachocline, at the base of the convection zone, the rotation period at various latitudes of the star tends to converge approaching the rotation rate at about 30° latitude at the surface, as in the case of the Sun (R. Howe et al. 2000).

The sample in this work cannot be simply analogized to the Sun, as the existence of the Kraft break implies significant structural differences between the two. Nevertheless, we believe that differential rotation would have a weaker impact within our sample than the Sun, because research on the Sun indicates that differential rotation primarily occurs in the surface convection zone, while stars above the Kraft break precisely lack a prominent surface convection zone. A larger error will lead to greater uncertainty in the slope during the correlation analysis. Therefore, we believe that it is better to overestimate the error as much as possible to avoid obtaining spurious results. Hence, we introduce an additional normalized error of 1.4–1 = 0.4. If the data were linearly distributed, this error would be significant. However, in reality, $RR_P$ spans multiple orders of magnitude, so this error does not have a strong impact. The final total error is then calculated as $e_{\rm final} = \sqrt{e^2 + (RR_P \cdot 0.28)^2 + (RR_P \cdot 0.4)^2}$, with the upper and lower errors calculated separately. The final total error was used in all subsequent analyses.

### 3.4. Analysis of Sample Membership in Galactic Substructures

For stars accreted by the Milky Way, their formation environments differ significantly from those of field stars in the Milky Way. If the samples originate from outside the Milky Way, more explanations can be considered. A star cluster in or near the Galaxy initially exhibited clustering properties. Through interaction with the Galaxy, it gradually disintegrates and forms a stellar stream (R. A. Ibata et al. 2002; K. V. Johnston et al. 2002; R. G. Carlberg 2012; R. Ibata et al. 2021); subsequently, it loses spatial distribution characteristics and retains only kinematic features, such as the Gaia sausage (V. Belokurov et al. 2018). Based on the proper motion, parallax, and radial velocity provided by Gaia DR3 (Gaia Collaboration et al. 2016, 2023), Galactic space-velocity components $U$, $V$, $W$; the velocities in the $x$, $y$, $z$ directions under the galactic coordinate system (D. R. H. Johnson & D. R. Soderblom 1987); and orbital elements, radius ($a$) and eccentricity ratio ($e$), under the McMillan17 gravitational potential (P. J. McMillan 2017), are calculated. The results in Figure 9 indicate that the samples lack distinct kinematic characteristics.

There are several young open clusters in the sample. Cross-referencing with T. Cantat-Gaudin & F. Anders (2020) reveals that, except for the previously mentioned NGC 6819, all the other young open clusters in the sample have ages less than 1 Gyr, totaling 163 stars (including NGC 6819). This number is negligible within the target sample. Thus the target samples can be considered to be field stars within the Galactic disk.

There is also a distinction between field stars within the Galactic disk, regarding whether they are located on spiral arms or far from them. Based on Y. Xu et al. (2023), we find that our sample involves two spiral arms, namely the Local Arm and the Perseus Arm, as illustrated in Figure 10. It should be noted that the coordinate system ($X$, $Y$, $Z$) used in Figure 10 is consistent with that in Y. Xu et al. (2023), but it differs from the system ($x$, $y$, $z$) used to calculate kinematic characteristics in Figure 9. The transformation relationship between them is $X = y$, $Y = 8.15 - x$, and $Z = z$.

For this particular sample, it is not feasible to analyze the influence of spiral arms on the rotational angular momentum.

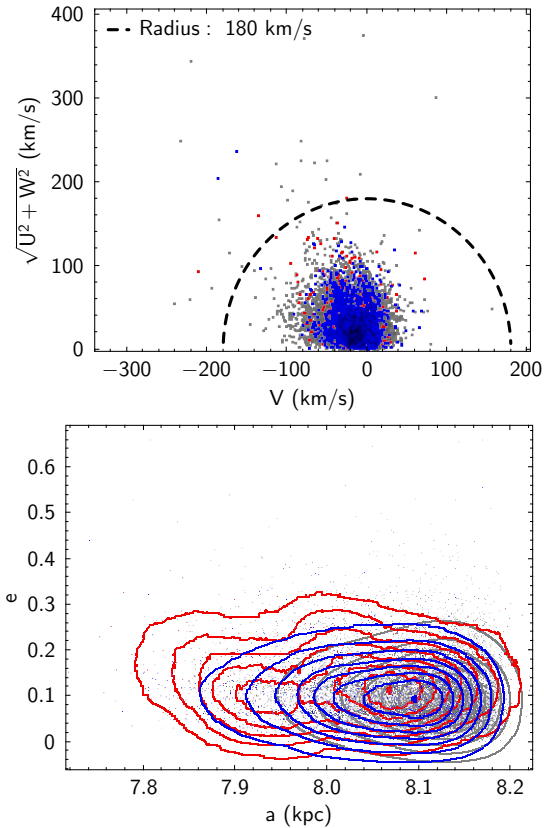

**Figure 9.** The distribution of samples in the kinematic parameter space. The dashed line in the top figure outlines the approximate range of the Galactic disk. The colors have the same meanings as in Figure 5.

This is due to the strong selection effect: within the line-of-sight direction of the Kepler field, samples that are closer to the spiral arms are farther from the observer, and there are fewer such samples. The vast majority of our sample is concentrated in the solar neighborhood and is also at a certain distance from the Local Arm. Therefore, we have neglected the impact of spiral arms.

### 3.5. Variation of Angular Momentum

First, we consider whether the theory of star formation can account for the excessively large angular momentum observed in some of the target samples. Then we consider the impact of the evolution of the angular momentum of the target samples themselves.

#### 3.5.1. Star–Disk Interaction during the PMS

The angular momentum of the target samples is 2–3 orders of magnitude higher than that of low-mass stars, prompting us to first consider the source of their initial angular momentum. The initial angular momentum of stars is inherited from molecular clouds but undergoes significant removal during star–disk interaction in the PMS. According to A. L. Rosen et al. (2012), for stars of the same mass, the varying parameters (their Section 4.2) have a minor impact on their radii, but the varying parameters significantly influence the stellar rotation period, which could reach 2–3 orders of magnitude. This indicates that by adjusting parameters during the PMS phase, such as the properties of molecular clouds, the observed

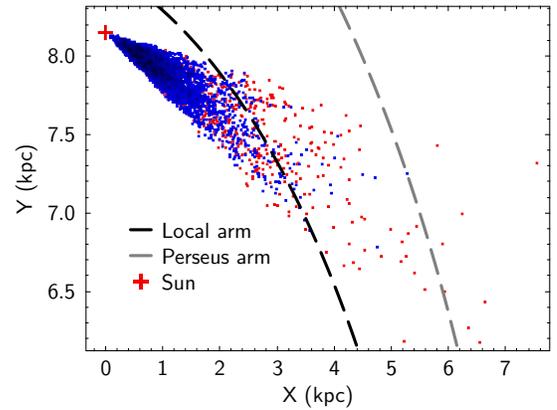

**Figure 10.** The distribution of the target sample in the galactic coordinate system, with the Local Arm marked by a dark dashed line and the Perseus Arm marked by a light dashed line. The locations of the arms are given by Y. Xu et al. (2023). The colors of the data points carry the same meanings as those in Figure 5.

angular momentum distribution, only in terms of absolute values, can be explained.

#### 3.5.2. Variation of Angular Momentum during the Main Sequence

It is widely acknowledged that stars above the Kraft break experience relatively minimal angular momentum loss owing to stellar winds, resulting in negligible magnetic spin-down during the main sequence phase. Furthermore, even if these stars undergo magnetic braking effects during turnoff phase, the duration of this evolutionary stage is insufficient for magnetic braking to exert a substantial influence.

### 3.6. Results of Analysis

Based on the stellar parameters (including rotation period) of 712 turnoff stars and 2982 MS stars, we used $RR_P$ to denote the angular momentum of stars. We observed that turnoff and MS stars of the same mass exhibit conservation of angular momentum and follow a similar angular momentum–age relationship. We ruled out the possibility that the observed relationship is due to selection effects. After accounting for the influence of the internal mass and velocity distributions of stars on $RR_P$, and verifying that the vast majority of the sample stars belong to the Galactic disk field population, we determined that the absolute values of observed angular momentum can be explained by the evolutionary theory of the PMS phase, but the relationship between angular momentum and age needs to be explained. Considering that angular momentum dissipation is weak for the sample stars during both the MS and PMS phases, we conclude that the angular momentum–age relationship of the sample reflects the variations in the star-forming environments over the history of the Milky Way.

## 4. Discussion

The observed angular momentum–age relation warrants further analysis to provide quantitative constraints on the evolutionary history of the Milky Way. Given that the age parameter appears to be influenced by mass-based priors, coupled with the fact that stars of different masses exhibit distinct age distribution ranges and internal mass distributions, dividing the sample into several mass categories can mitigate a series of confounding effects. We divide the target sample into four populations: 1.2–1.4$M_\odot$, 1.4–1.6$M_\odot$, 1.6–1.8$M_\odot$, and >1.8$M_\odot$. Subsequently, we will analyze these four populations using the same analytical method.

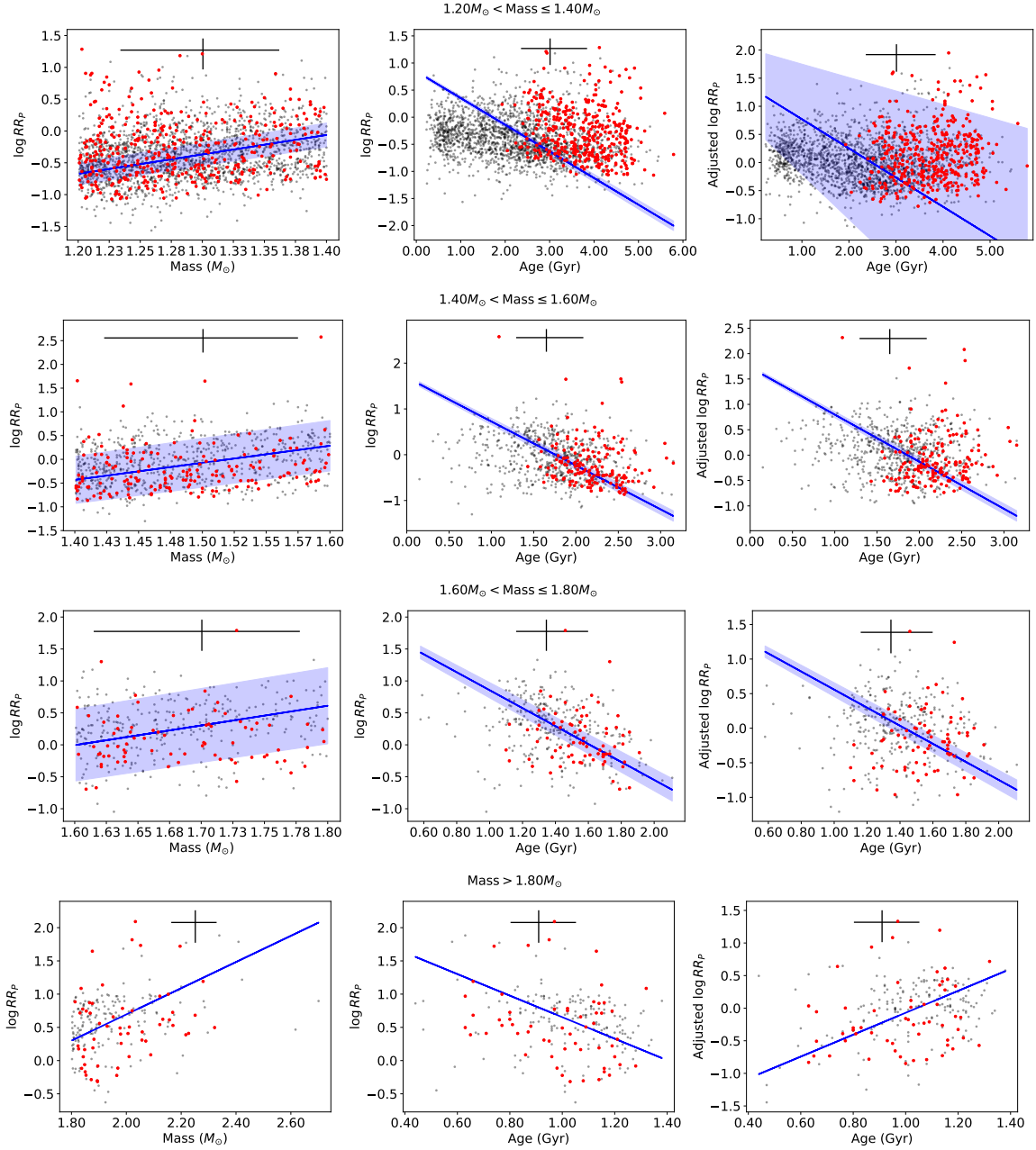

**Figure 11.** Each row of panels shares the same mass range, specifically 1.2–1.4$M_\odot$, 1.4–1.6$M_\odot$, 1.6–1.8$M_\odot$, and >1.8$M_\odot$, respectively. The star numbers are 2292, 799, 373, and 228. Each column of subplots presents the same set of axes: $RR_P$–mass, $RR_P$–age, and adjusted $RR_P$–age. This paper outlines a method for calculating the false-positive rate based on these uncertainties. The crosshair symbol represents a typical error, corresponding to the median of the errors. Turnoff stars are marked with red circles, numbering 409, 169, 77, and 57. The fitting process incorporates both the turnoff stars and the MS stars. The straight blue line represents the fitted line, and the light-blue area indicates the reference range provided by the bootstrap method. The reference range for the fitted line is not shown in the fourth row because its fitted values are meaningless, so bootstrap was not carried out.

Our analysis consists of four steps. In the first step, we perform a linear fit of log $RR_P$ versus mass. In the second step, we conduct a linear fit of log $RR_P$ versus age. In the third step, we subtract the straight line obtained from the log $RR_P$–mass linear fit from log $RR_P$ to derive the adjusted log $RR_P$–age, and then perform a linear fit on this adjusted value. The reason for using adjusted log $RR_P$–age is to ensure that the resulting angular momentum–age relationship is not influenced by mass.

After all, stars with greater masses tend to have a narrower age distribution range and possess larger angular momenta; even after dividing the sample into several small mass intervals, it is still challenging to guarantee that this effect is entirely eliminated. By comparing the fitting results of log $RR_P$–age with those of adjusted log $RR_P$–age, we can determine whether the observed linear relationship is real or merely an artifact caused by a mass–age relationship.

It is relatively easy to determine whether a correlation exists. However, to obtain a slope with reference value during linear fitting, it is essential to account for parameter errors, and the upper and lower errors must be considered separately

Table 1
Fitting Results for $RR_P$ Relationships

| Mass Range ($M_\odot$) (1) | Relation (2) | Value | | | | | | | |
|---|---|---|---|---|---|---|---|---|---|
| | | Slope (3) | Slope Err (4) | FPR (5) | Intercept (6) | Int. Err (7) | Tot. Scat. (8) | Err. Scat. (9) | Int. Scat. (10) |
| 1.2–1.4 | $\log RR_P$ vs. $M_*$ | 3.05 | 0.07 | 0.00 | −4.33 | 0.09 | 0.40 | 0.06 | 0.39 |
| | $\log RR_P$ vs. age | −0.49 | 0.01 | 0.00 | 0.84 | 0.03 | 1.40 | 0.63 | 1.25 |
| | Adj. $\log RR_P$ vs. age | −0.52 | 0.28 | 0.06 | 1.29 | 0.70 | 1.50 | 0.63 | 1.36 |
| 1.4–1.6 | $\log RR_P$ vs. $M_*$ | 3.59 | 0.17 | 0.00 | −5.47 | 0.26 | 0.45 | 0.06 | 0.45 |
| | $\log RR_P$ vs. age | −0.96 | 0.02 | 0.00 | 1.68 | 0.05 | 0.76 | 0.38 | 0.66 |
| | Adj. $\log RR_P$ vs. age | −0.93 | 0.02 | 0.00 | 1.73 | 0.04 | 0.84 | 0.38 | 0.75 |
| 1.6–1.8 | $\log RR_P$ vs. $M_*$ | 3.08 | 0.17 | 0.00 | −4.94 | 0.29 | 0.44 | 0.06 | 0.43 |
| | $\log RR_P$ vs. age | −1.40 | 0.05 | 0.00 | 2.26 | 0.07 | 0.52 | 0.16 | 0.50 |
| | Adj. $\log RR_P$ vs. age | −1.31 | 0.04 | 0.00 | 1.86 | 0.06 | 0.61 | 0.16 | 0.59 |
| >1.8 | $\log RR_P$ vs. $M_*$ | 1.97 | ⋯ | ⋯ | −3.24 | 0.33 | ⋯ | ⋯ | ⋯ |
| | $\log RR_P$ vs. age | −1.62 | ⋯ | ⋯ | 2.27 | 0.10 | ⋯ | ⋯ | ⋯ |
| | Adj. $\log RR_P$ vs. age | 1.68 | ⋯ | ⋯ | −1.75 | 0.06 | ⋯ | ⋯ | ⋯ |

**Note.** This table shows fitting results for $RR_P$ relationships across different stellar mass ranges. The three relationships examined are $\log RR_P$ vs. stellar mass ($M_*$), $\log RR_P$ vs. age, and adjusted $\log RR_P$ vs. age. For each relationship, we report the slope, slope error, false-positive rate (FPR), intercept, intercept error, total scatter, error scatter, and intrinsic scatter. Mass ranges are in units of $M_\odot$.

because the upper and lower errors of stellar ages can differ significantly. Additionally, Galactic field stars of the same age may have experienced different star-forming environments. Moreover, even when the star-forming environment and stellar mass are identical, there can still be a distribution of angular momenta at the ZAMS. Consequently, an intrinsic scatter should exist in the angular momentum–age relationship. In summary, our linear fitting needs to simultaneously consider the upper and lower errors of both variables as well as the intrinsic scatter of the data points. To this end, we have improved the commonly used fitting method and named it the *least pseudoorthogonal weighted residual fitting method*, which is described in detail in Appendix B.

Based on the aforementioned methods, the results for the mass ranges of 1.2–1.4$M_\odot$, 1.4–1.6$M_\odot$, 1.6–1.8$M_\odot$, and >1.8$M_\odot$ are shown in Figure 11. Here, the adjusted $RR_P$ is derived by subtracting the fitted line of $RR_P$–mass from $RR_P$, aiming to eliminate the influence of the mass–age relationship on $RR_P$–age. Two stars with masses greater than 4$M_\odot$ were discarded as outliers, and these two stars do not affect any of the conclusions because, as will be mentioned later, the fitting results above 1.8$M_\odot$ are not acceptable. Table 1 lists all the parameters in Figure 11. If the false-positive rates are larger than 0.05, the fitted relationship is deemed nonexistent. For the 1.2–1.4$M_\odot$ mass range, the false-positive rate of the slope is 0.06, and thus the hypothesis of an angular momentum–age relationship is not accepted. For the 1.4–1.6$M_\odot$ and 1.6–1.8$M_\odot$ mass ranges, the false-positive rates of the slopes are both very close to 0, indicating with near certainty the existence of an angular momentum–age relationship for stars within these mass ranges. As previously mentioned, selection effects in the sample can lead to a spurious positive correlation, so the actual absolute value of the slope may be even larger. The slope for the >1.8$M_\odot$ mass range is not meaningful because, as shown in Figure 11, after adjusting $\log RR_P$–age based on $\log RR_P$–mass, the slope changes from negative to positive, suggesting that the previous negative slope was an artifact caused by the short life-span of massive stars. In Figure 11, the mass errors are relatively large compared to the selected mass range. However, this does not affect the conclusions because the errors have already been propagated to the age errors. The direct impact of the mass errors is that some stars below the Kraft break are mixed into the 1.2–1.4$M_\odot$ sample. This is probably the reason why the angular momentum–age relationship for the 1.2–1.4$M_\odot$ mass range is not accepted. Moreover, because of the relatively large mass errors, this work is unable to estimate the mass corresponding to the Kraft break.

In summary, stars with masses in the 1.4–1.6$M_\odot$ and 1.6–1.8$M_\odot$ ranges exhibit a clear angular momentum–age relationship, with angular momentum decreasing with age. We believe that this angular momentum–age relationship reflects changes in the star-forming environment over the history of the Milky Way. There might be two reasons why no correlation is observed for stars in the 1.2–1.4 $M_\odot$ (solar mass) range: First, these stars could be a mixture of those below the Kraft break. Due to the self-regulating nature of magnetic stellar winds, the angular momentum converges to a mass-dependent specific value, and thus it cannot reflect the characteristics of the molecular cloud. Second, these stars may have undergone strong braking during the PMS phase, where magnetic braking is the dominant mechanism, and this process is also self-regulating.

Considering the intrinsic scatters present in the fits for stars in the mass ranges of 1.2–1.4$M_\odot$, 1.4–1.6$M_\odot$, and 1.6–1.8$M_\odot$, this suggests that even field stars of the same age may have formed in different environments. We further analyzed the [Fe/H]–residual relationships for these three samples, as shown in Figure 12. Table 2 lists all the parameters presented in Figure 12. The false-positive rates for the [Fe/H]–residual correlations are 0.24, 0.03, and 0.25 for the 1.2–1.4$M_\odot$, 1.4–1.6$M_\odot$, and 1.6–1.8$M_\odot$ samples, respectively. The relatively low false-positive rate for the 1.4–1.6$M_\odot$ sample is due to the presence of one star with [Fe/H] less than −1. If this star is excluded from the analysis, the false-positive rate also exceeds 0.2. Of course, this also implies that if the [Fe/H] distribution were broader, a correlation might likely be observed. However, given that there is only one star with [Fe/H] less than −1 in our current sample, no reliable conclusion can be drawn, highlighting the necessity of expanding the sample in future studies. Based on the current data, [Fe/H] is not the cause of the observed intrinsic scatter.

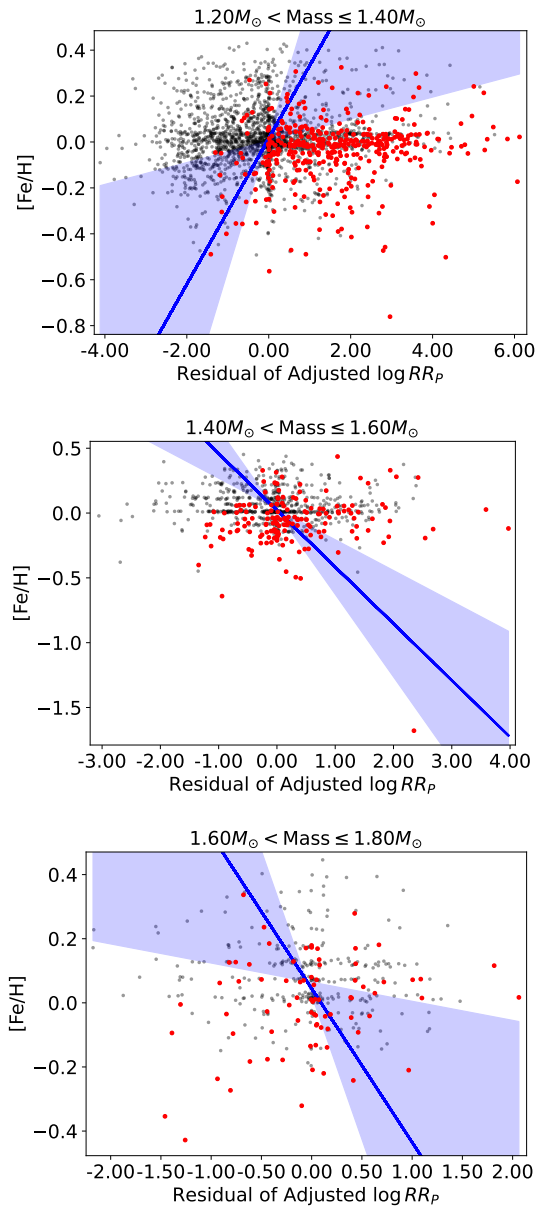

**Figure 12.** The mass ranges of the samples are provided above each of the three panels. The meanings of various symbols are the same as those in Figure 11. The correlations in all three plots are deemed unacceptable. In the second panel, there is a metal-poor star that results in a relatively small bootstrap error; if that star is removed, the linear relationship still cannot be accepted.

Considering that stars with masses of 1.2–1.4$M_\odot$ are still observable in today's young open clusters but those with masses greater than 1.4$M_\odot$ are extremely rare, combined with the phenomena found in this work, the following explanation can be given: Within the mass range of approximately 1.2–2$M_\odot$, as the mass increases, the stars become more sensitive to the formation environment. By "sensitive," we mean that for a given variation in molecular cloud parameters, more sensitive stars will exhibit larger changes in angular momentum at the ZAMS phase. Within the range of variations in molecular cloud parameters throughout the history of the Milky Way, 1.4$M_\odot$ can serve as a vague boundary; stars below this mass and above the Kraft break exhibit no significant change in angular momentum at the ZAMS, while those above this mass exhibit significant changes, to the extent that an angular momentum–age relationship exists for stars above this mass. Molecular clouds that have had star formation recently tend to form stars with greater angular momentum, but excessively large angular momentum can promote fragmentation and inhibit star formation, making it increasingly difficult to form more massive stars today. This also partially explains the initial mass function (IMF) problem. Several observations and theories suggest that the IMF slope for massive stars in globular clusters (GCs) depends on the initial cloud density and metallicity (Z). The IMF becomes increasingly top-heavy with decreasing Z and increasing gas density (e.g., M. Marks et al. 2012). It can then be conjectured that in the history of the Milky Way, the abundances of molecular clouds gradually increased with age, while their density gradually decreased, leading to the formation of stars with masses greater than 1.4 solar masses having higher angular momentum. Today, this angular momentum has become so high that it severely inhibits the formation of stars with masses greater than 1.4 solar masses. The physical mechanisms mentioned here should be of great importance for stars with masses in the range of 1.4–2$M_\odot$. As for stars with masses above 2$M_\odot$, the primary inhibiting factor is that the radiative pressure from the protostar prevents further mass accretion, and this is an effect that strongly depends on metallicity. In fact, the abundance–age relationship of molecular clouds in the history of the Milky Way in this conjecture has been confirmed by the stellar metallicity–age relationship obtained from large-sample spectroscopic surveys such as LAMOST (X.-Q. Cui et al. 2012; G. Zhao et al. 2012; X.-W. Liu et al. 2015; H. Yan et al. 2022). The increase in metallicity over time is a generally known fact, simply by stars slowly enriching the Galaxy. This closed logical loop further demonstrates the reliability of this series of conjectures. In future work, more precise numerical simulations of the PMS phase are needed to further clarify the evolutionary history of the star-forming environment in the Milky Way.

## 5. Conclusion

This work posits that the rotational angular momenta of MS and turnoff stars above the Kraft break reflects the evolutionary history of the Milky Way. The sample primarily consists of field stars in the solar neighborhood. The angular momentum per unit mass of these stars is mainly correlated with stellar mass, followed by age. No clear influence of [Fe/H] on angular momentum has been observed. This is partly due to the narrow [Fe/H] distribution range of the sample and partly because [Fe/H] only reflects one aspect of the star-forming environment. After accounting for measurement errors and those arising from the internal mass distribution and rotational velocity distribution of the stars, we find that the 1.2–1.4$M_\odot$ sample does not exhibit a clear angular momentum–age relationship, while the 1.4–1.6$M_\odot$ and 1.6–1.8$M_\odot$ samples exhibit clear negative correlations. We suggest that stars in the 1.2–1.4$M_\odot$ range may be mixed with stars lower than Kraft break, or they should experience significant angular momentum loss during the PMS phase. The more massive

Table 2
Fitting Results between [Fe/H] and Residuals of Adjusted log $RR_P$

| Mass Range ($M_\odot$) (1) | Relation (2) | Value | | | | | | | |
|---|---|---|---|---|---|---|---|---|---|
| | | Slope (3) | Slope Err (4) | FPR (5) | Intercept (6) | Int. Err (7) | Tot. Scat. (8) | Err. Scat. (9) | Int. Scat. (10) |
| 1.2–1.4 | [Fe/H] versus Res. of Adj. log $RR_P$ | 0.32 | 0.27 | 0.24 | 0.02 | 0.01 | 1.62 | 0.03 | 1.62 |
| 1.4–1.6 | [Fe/H] versus Res. of Adj. log $RR_P$ | −0.44 | 0.20 | 0.03* | 0.02 | 0.01 | 0.95 | 0.04 | 0.95 |
| 1.6–1.8 | [Fe/H] versus Res. of Adj. log $RR_P$ | −0.48 | 0.42 | 0.25 | 0.04 | 0.02 | 0.72 | 0.04 | 0.72 |

**Note.** *There is a metal-poor star that results in a relatively small bootstrap error; if that star is removed, the linear relationship still cannot be accepted. This table shows fitting results for [Fe/H] relationships across different stellar mass ranges. The parameters are the same as Table 1.

stars do not dissipate enough angular momentum during the PMS phase, thus retaining more information about their star-forming environments. For the 1.4–1.6$M_\odot$ and 1.6–1.8$M_\odot$ samples, we provide the slopes and their error ranges. We also find that the scatter caused by measurement errors cannot account for the observed scatter in the angular momentum–age distribution of the samples. This intrinsic scatter may have multiple explanations, potentially suggesting the presence of unknown systematic biases or that the star-forming environments at the same time in the Milky Way's history were different. We demonstrate that this intrinsic scatter is independent of [Fe/H], which largely indicates that within the range of −0.5 < [Fe/H] < 0.5, differences in stellar rotational angular momenta are not primarily caused by the [Fe/H] of their formation environments. However, this does not rule out the possibility of observing correlations if the [Fe/H] range is expanded.

Combining the findings of this work with past discoveries, we propose a nonexclusive but self-consistent explanation: As the Milky Way evolves, at least within the past 6 Gyr, its environment has increasingly favored the formation of stars with greater rotational angular momenta. The rotational angular momenta of stars in the 1.4–1.8$M_\odot$ range retain information about their formation environments, thus serving as "witnesses" to this evolutionary history. This trend in environmental changes suppresses the formation of high-mass stars, as excessively rapid rotation during the PMS phase can promote fragmentation, with more massive stars being more severely affected, potentially leading to changes in the IMF.

In the future, theoretical efforts should refine numerical simulations of the PMS phase to distinguish differences in the PMS phase among stars within the 1.2–2$M_\odot$ range. Observationally, there is a need to expand the [Fe/H] distribution of the stellar sample above the Kraft break.

## Acknowledgments

This work has made use of data from the European Space Agency (ESA) mission Gaia (https://www.cosmos.esa.int/gaia), processed by the Gaia Data Processing and Analysis Consortium (DPAC; https://www.cosmos.esa.int/web/gaia/dpac/consortium). Funding for the DPAC has been provided by national institutions, in particular the institutions participating in the Gaia Multilateral Agreement.

This paper includes data collected by the Kepler mission, specifically the KeplerKIC repository (STScI). Funding for the Kepler mission is provided by the NASA Science Mission Directorate. STScI is operated by the Association of Universities for Research in Astronomy, Inc., under NASA contract NAS 5–26555.

This work is supported by funds from the first author's institution.

The authors sincerely thank the anonymous reviewers for their constructive comments and valuable suggestions, which significantly contributed to improving the quality of this manuscript. The authors also thank the editors for their kind assistance throughout the peer-review process.

*Software:* TOPCAT (M. B. Taylor 2005), astropy (Astropy Collaboration et al. 2013, 2018, 2022), numpy, scipy (P. Virtanen et al. 2020).

## Author Contributions

Y.-F.S. was responsible for conceptualization, data curation, formal analysis, methodology, and writing the original draft.

Y.X. was responsible for supervision and project administration.

Y.-B.W., X.-L.H., X.-X.H., and Q.Y. were responsible for the review and editing.

## Appendix A
## Effect of $I/MR^2$

A. Claret & A. Gimenez (1989) define

$$\beta = (I/MR^2)^{1/2}$$

and provide a grid of $\beta$ values for stars of different masses and ages. However, the parameter range covered by the provided grid points is limited. As a result, the $\beta$ values obtained through interpolation only encompass a portion of the sample. Figure 13 illustrates the relationship between $\beta$ and stellar mass/age, as well as the distribution of the sample before and after multiplying $RR_P$ by $\beta^2$. It can be observed that the inherent limitations in the parameter range of $\beta$ already introduce an artificial correlation in the sample selection. This artificial correlation becomes even more pronounced after multiplying by $\beta^2$. Nevertheless, the relative positions of the data points are not significantly affected before and after multiplying by $\beta^2$, because as previously discussed, the variations in $\beta$ are relatively small compared to the orders-of-magnitude differences in rotational angular momentum.

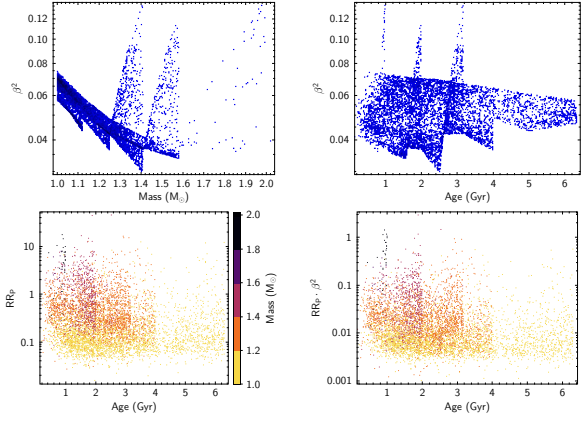

Figure 13. Mass-$\beta$ distribution, age-$\beta$ distribution of samples capable of undergoing $\beta$ interpolation, as well as the age–$RR_P$ distributions before and after multiplying by $\beta^2$. In the panels on the second row, colors represent mass.

## Appendix B
## The Least Pseudoorthogonal Weighted Residual Fitting Method

First, we define the Gaussian weight. The Gaussian weight of a point $x$ with an error $e_x$ from point $a$ is given by

$$W_G\left(\frac{|x-a|}{e_x}\right) = \Phi\left(\frac{|x-a|}{e_x}\right) - \Phi\left(-\frac{|x-a|}{e_x}\right).$$

When $x > a$, the lower error of $e_x$ is used; when $x < a$, the upper error of $e_x$ is used:

$$\Phi(z) = \int_{-\infty}^{z} \frac{1}{\sqrt{2\pi}} e^{-t^2/2} dt.$$

The Gaussian weight has the following properties: $W_G(k) = \{0,$ when $\frac{|x-a|}{e_x} = 0$, $0.683$, when $\frac{|x-a|}{e_x} = 1$, $0.955$, when $\frac{|x-a|}{e_x} = 2$, $0.997$, when $\frac{|x-a|}{e_x} = 3\}$.

For any parameters $a$ and $b$, the fitted line is $y = ax + b$. For a point $\{x_i, y_i\}$, the pseudoorthogonal weighted residual is

$$\mathrm{Res}_i^2 = W_G\left(\frac{|\Delta X_i|}{e_{x_i}}\right)^2 \cdot \Delta X_i^2 + W_G\left(\frac{|\Delta Y_i|}{e_{y_i}}\right)^2 \cdot \Delta Y_i^2,$$

where $\Delta X_i = x_i - (y_i - b)/a$ and $\Delta Y_i = y_i - (ax_i + b)$. The parameters $a$ and $b$ are determined by minimizing $\sum_i \mathrm{Res}_i$. $e_{x_i}$ and $e_{y_i}$ are upper error or lower error, depending on the relative position of point $\{x_i, y_i\}$ and line $y = ax + b$. Then, the bootstrap method is used to estimate the error range of the parameters. In this work, the error is set as the standard deviation of the parameters obtained from 100 bootstrap iterations. The false-positive rate of the slope can be represented by

$$1 - W_G\left(\frac{|\mathrm{slope}|}{\mathrm{slope}_{\mathrm{err}}}\right).$$

Total scatter is defined as

$$\sigma_{\mathrm{total}} = \sigma(\{Res_i\}).$$

Error scatter is defined as

$$\sigma_{\mathrm{error}}^2 = \sigma^2(\{x_{e+}\}) + \sigma^2(\{x_{e-}\}) + \sigma^2(\{y_{e+}\}) + \sigma^2(\{y_{e-}\}).$$

Intrinsic scatter is defined as

$$\sigma_{\mathrm{int}}^2 = \sigma_{\mathrm{total}}^2 - \sigma_{\mathrm{error}}^2.$$

The code has been made publicly available in Zenodo (Y.-F. Shen 2025). A working copy is also in github at https://github.com/nickname970514/least-pseudo-orthogonal-weighted-residual-fitting/.

## Appendix C
## Data Catalog

This section provides the raw data involved in the construction of Table 1, as well as all the raw data used in this work (see Table 3).

Table 3
Catalog

| Column | Format | Units | Label | Description |
|---|---|---|---|---|
| 1 | I8 | … | KID | Kepler ID |
| 2 | F6.3 | d | PRot | A. McQuillan et al. (2014) rotation period |
| 3 | F6.3 | d | e_PRot | Uncertainty in PRot |
| 4 | F9.5 | solRad2/d | RRP | Stellar radius squared divided by rotation period |
| 5 | E8.2 | solRad2/d | E_RRP | Upper uncertainty in RRP |
| 6 | E8.2 | solRad2/d | e_RRP | Lower uncertainty in RRP |
| 7 | E8.2 | solRad2/d | upRRfinal | Final upper uncertainty in RRP; see Section 3.3 |
| 8 | E8.2 | solRad2/d | loRRfinal | Final lower uncertainty in RRP; see Section 3.3 |
| 9 | F11.7 | deg | RAdeg | R.A. in decimal degrees (J2000) |
| 10 | F11.8 | deg | DEdeg | decl. in decimal degrees (J2000) |
| 11 | E8.2 | … | Pbin | Probability of being a binary; Gaia |
| 12 | F7.4 | solRad | RFl | Gaia FLAME stellar radius |
| 13 | F7.4 | solRad | b_RFl | Lower boundary on RFl |
| 14 | F7.4 | solRad | B_RFl | Upper boundary on RFl |
| 15 | F5.3 | solMass | MFl | Gaia FLAME stellar mass |
| 16 | F5.3 | solMass | b_MFl | Lower boundary on MFl |
| 17 | F5.3 | solMass | B_MFl | Upper boundary on MFl |
| 18 | F6.3 | Gyr | AgeFl | Gaia FLAME stellar age |
| 19 | F6.3 | Gyr | b_AgeFl | Lower boundary on AgeFl |
| 20 | F6.3 | Gyr | B_AgeFl | Upper boundary on AgeFl |
| 21 | A28 | … | Gaia | Gaia identifier |
| 22 | F7.4 | mas | Plx | Gaia DR3 parallax |
| 23 | F6.4 | mas | e_Plx | Uncertainty in Plx |
| 24 | F7.3 | arcsec/yr | pmRA | Gaia DR3 proper motion along RA |
| 25 | F5.3 | arcsec/yr | e_pmRA | Uncertainty in pmRA |
| 26 | F7.3 | arcsec/yr | pmDE | Gaia DR3 proper motion along DE |
| 27 | F5.3 | arcsec/yr | e_pmDE | Uncertainty in pmDE |
| 28 | F6.3 | … | RUWE | Gaia DR3 Re-normalized Unit Weight Error |
| 29 | F7.2 | km/s | RVel | Gaia DR3 radial velocity |
| 30 | F5.2 | km/s | e_RVel | Uncertainty in RVel |
| 31 | F6.2 | solRad | $a$ | Orbital radius within the Milky Way from galpy.orbit.Orbit.r() |
| 32 | F4.2 | … | $e$ | Orbital ellipticity within the Milky Way from galpy.orbit.Orbit.e(McMillan17) |
| 33 | F5.2 | kpc | $x$ | Galactic $X$ position from SkyCoord().galactic.cartesian.x |
| 34 | F6.2 | kpc | $y$ | Galactic $X$ position from SkyCoord().galactic.cartesian.y |
| 35 | F5.2 | kpc | $z$ | Galactic $X$ position from SkyCoord().galactic.cartesian.z |
| 36 | F7.2 | km/s | $U$ | Galactic $U$ space-velocity component |
| 37 | F7.2 | km/s | $V$ | Galactic $V$ space-velocity component |
| 38 | F7.2 | km/s | $W$ | Galactic $W$ space-velocity component |
| 39 | F5.3 | solMass | MassB20 | T. A. Berger et al. (2020) stellar mass |
| 40 | F5.3 | solMass | E_MassB20 | Upper uncertainty in MassB20 |
| 41 | F5.3 | solMass | e_MassB20 | Lower uncertainty in MassB20 |
| 42 | F7.1 | K | TeffB20 | T. A. Berger et al. (2020) stellar effective temperature |
| 43 | F6.1 | K | E_TeffB20 | Upper uncertainty in TeffB20 |
| 44 | F6.1 | K | e_TeffB20 | Lower uncertainty in TeffB20 |
| 45 | F5.3 | [cm/s2] | loggB20 | T. A. Berger et al. (2020) stellar log surface gravity |
| 46 | F5.3 | [cm/s2] | E_loggB20 | Upper uncertainty in loggB20 |
| 47 | F5.3 | [cm/s2] | e_loggB20 | Lower uncertainty in loggB20 |
| 48 | F6.3 | [-] | [Fe/H]B20 | T. A. Berger et al. (2020) stellar metallicity |
| 49 | F5.3 | [-] | E_[Fe/H]B20 | Upper uncertainty in [Fe/H]B20 |
| 50 | F5.3 | [-] | e_[Fe/H]B20 | Lower uncertainty in [Fe/H]B20 |
| 51 | F6.3 | solRad | RadB20 | T. A. Berger et al. (2020) stellar radius |
| 52 | F5.3 | solRad | E_RadB20 | Upper uncertainty in RadB20 |
| 53 | F5.3 | solRad | e_RadB20 | Lower uncertainty in RadB20 |
| 54 | F5.2 | Gyr | AgeB20 | T. A. Berger et al. (2020) stellar age |
| 55 | F5.2 | Gyr | E_AgeB20 | Upper uncertainty in AgeB20 |
| 56 | F5.2 | Gyr | e_AgeB20 | Lower uncertainty in AgeB20 |
| 57 | A5 | … | AboveKB | Source has RUWE < 1.4 and a mass > 1.2 solar |
| 58 | A5 | … | Turnoff | Source is AboveKB and at the main sequence turnoff |
| 59 | A5 | … | MS | Source is AboveKB and on the main sequence |
| 60 | F9.6 | mag | Gmag | Gaia DR3 $G$ band apparent magnitude |
| 61 | F8.6 | mag | e_Gmag | Uncertainty in Gmag |

**Note.** Table 3 is published in its entirety in the electronic edition of the *Astrophysical Journal*. A portion is shown here for guidance regarding its form and content.
(This table is available in its entirety in machine-readable form in the online article.)

## ORCID iDs

Yu-Fu Shen (申淯夫)  https://orcid.org/0000-0003-4445-6504